\documentclass[12pt]{article}
\usepackage{epsfig}
\usepackage{a4,isolatin1}
\usepackage{amsmath,amsfonts,latexsym, amssymb}

\newtheorem{satz}{Theorem}[section]
\newtheorem{defi}[satz]{Definition}

\newtheorem{bem}[satz]{Remark}
\newtheorem{lemma}[satz]{Lemma}
\newtheorem{koro}[satz]{Corollary}
\newtheorem{bsp}[satz]{Example}
\newtheorem{assumption}[satz]{Assumption}

\newtheorem{conclusion}[satz]{Conclusion}
\newtheorem{ob}[satz]{Observation}

\newtheorem{problem}[satz]{Problem}

\newcommand{\mcal}{\mathcal}

\newcommand{\tit}{\textit}

\newcommand{\C}{\mathbb{C}}
\newcommand{\N}{\mathbb{N}}

\newcommand{\Z}{\mathbb{Z}}

\newcommand{\bewende}{$ \hfill \Box $}

\newcommand{\Fc}{\cal F}

\newcommand{\dast}{d^{\ast}}

\begin{document}
\thispagestyle{empty}
\begin{center}
\vspace*{1.0cm}

{\LARGE{\bf Dirac Operators  and the Calculation of the Connes
    Metric\\on arbitrary  (Infinite) Graphs}}

\vskip 1.5cm

{\large {\bf Manfred Requardt }}

\vskip 0.5 cm

Institut f\"ur Theoretische Physik \\
Universit\"at G\"ottingen \\
Bunsenstrasse 9 \\
37073 G\"ottingen \quad Germany\\
(E-mail: requardt@theorie.physik.uni-goettingen.de)

\end{center}

\vspace{1 cm}

\begin{abstract}
  As an outgrowth of our investigation of non-regular spaces within
  the context of quantum gravity and non-commutative geometry, we
  develop a graph Hilbert space framework on arbitrary (infinite) graphs
  and use it to study spectral properties of \tit{graph-Laplacians} and
  \tit{graph-Dirac-operators}. We define a {\it spectral triplet}
  sharing most of the properties of what Connes calls a \tit{spectral
    triple}. With the help of this scheme we derive an explicit
  expression for the {\it Connes-distance function} on general
  directed or undirected graphs. We derive a series of apriori
  estimates and calculate it for a variety of examples of
  graphs. As a possibly interesting aside, we show that the natural
  setting of approaching such problems may be the framework of
  \tit{(non-)linear programming} or \tit{optimization}.  We compare
  our results (arrived at within our particular framework) with the
  results of other authors and show that the seeming differences
  depend on the
  use of different graph-geometries and/or Dirac operators.\\[0.5cm]
\end{abstract} \newpage
\setcounter{page}{1}

\section{Introduction}
In recent years we started a programme to reconstruct continuum
physics and/or mathematics from an underlying more primordial and
basically discrete theory living on the Planck-scale (cf.
\cite{1} to \cite{3} or \cite{requroy}). As sort of a
``spin-off'' various problems of a more mathematical and technical
flavor emerged which may have an interest of their own. \tit{Discrete
  differential geometric} concepts were dealt with in \cite{1}, the
theory of \tit{random graphs} was a central theme of \cite{2}, topics
of \tit{dimension theory} and \tit{fractal geometry} were addressed in
\cite{3}, \tit{lump-spaces} and \tit{random metrics} in \cite{requroy}.

If one wants to recover the usual (differential) operators (or
more generally, the concepts of standard functional analysis), being
in use in ordinary continuum physics and mathematics by some sort of
limiting process from their discrete protoforms, which live, on their
part, on a relatively disordered discrete background like, say, a
network, it is reasonable to analyse in a first step these discrete
counterparts more closely. This will be one of our themes in the
following with particular emphasis on discrete \tit{Laplace-} and
\tit{Dirac-operators} on general \tit{graphs}. In contrast to
\cite{graph} we now also include arbitrary \tit{directed graphs}. The
main thrust goes however into an analysis of metrical concepts on
discrete (non-commutative) spaces like general graphs, being induced
by \tit{graph Dirac-operators} and the \tit{Connes-distance functional}.

We note in passing that \tit{functional analysis on graphs} is both of
interest in pure and applied mathematics and also in various fields of
(mathematical) physics. For one, discrete systems have an increasing
interest of their own or serve as easier to analyse prototypes of
their continuum counterparts. To mention a few fields of applications:
\tit{graph theory} in general, analysis on \tit{(discrete) manifolds},
\tit{lattice} or \tit{discretized} versions of physical models in
statistical mechanics and quantum field theory, \tit{non-commutative
  geometry}, \tit{networks}, \tit{fractal geometry} etc. From the
widely scattered (mathematical and physical) literature we mention
(possibly) very few sources we are aware of. Some ot them were of
relevance for our own motivation, some others we came across
only recently (see e.g.  \cite{Harary} to \cite{Weaver}, \cite{4} and
\cite{Landi}, \cite{Iochum} or \cite{Novikov}). Some more literature
like e.g.  \cite{Davies} was pointed out to us by Mueller-Hoissen; the
possible relevance of references \cite{Colin} to \cite{Schrader} were
brought to our notice by some unknown referee. Last, but not least,
there is the vast field of \tit{discretized quantum gravity} (see e.g.
\cite{Loll} or \cite{Ambjorn}).  All this shows that the sort of
discrete functional analysis we are dealing with in the following, is
presently a very active field with a lot of different applications.

For the convenience of the reader we begin with compiling some
concepts and tools dealing with graph Hilbert spaces which we
then use to investigate the spectral properties of \tit{graph
  Laplacians} and \tit{Dirac operators}. In a next step we study and
test concepts and ideas, which arose in the context of
\tit{non-commutative geometry}. As we (and others) showed in preceding
papers, networks and graphs may (or even should) be understood as
examples of \tit{non-commutative} spaces. A currently interesting
topic in this field is the investigation of certain \tit{distance
  functionals} on ``nasty'' or \tit{non-standard} spaces and their
mathematical or physical ``naturalness''.  Graphs carry, on the one
hand, a natural {\it metric structure} given by a {\it distance
  function} $d(x,y)$, with $x,y$ two {\it nodes} of the graph (see the
following sections). This fact was already employed by us in e.g.
\cite{3} to develop dimensional concepts on graphs. Having Connes'
concept of distance in noncommutative geometry in mind (cf.  chapt. VI
of \cite{4}), it is natural to try to compute it in model systems,
which means in our context: arbitrary graphs, and compare it with the
already existing notion of graph distance mentioned above.  (We note
in passing that the calculation of the Connes distance for general
graphs turns out to be surprisingly complex and leads to perhaps
unexpected connections to fields of mathematics like e.g.
\tit{(non-)linear programming} or \tit{optimization}; see the last
section).

Therefore, as one of many possible applications of our formalism, we
construct a protoform of what Connes calls a {\it spectral triple},
that is, a Hilbert space structure , a corresponding representation of
a certain (function) algebra and a (in our framework) natural
candidate for a so-called {\it Dirac operator} (not to be confused
with the ordinary Dirac operator of the Dirac equation), which encodes
certain properties of the {\it graph geometry}. This will be done in
section \ref{Triplet}.

In the last (and central) section, which deals with the distance
concept deriving from this spectral triplet we will investigate this
concept more closely as far as graphs and similar spaces are
concerned. In this connection some recent work should be mentioned, in
which Connes' distance function was analyzed in certain simple models
like e.g. one-dimensional lattices (\cite{5}-\cite{7}). These papers
already show that it is a touchy business to isolate ``the''
appropriate Dirac operator (after all, different Dirac operators are
expected to lead to different geometries!)  and that it is perhaps
worthwhile to scrutinize the whole topic in a more systematic way. We
show in particular that one may choose different Dirac-operators on
graphs (or rather, different types of graphs over the same node set)
which may lead to different results for e.g. the corresponding
Connes-distance.

The problem of finding suitable metrics on ``non-standard'' spaces is
a particularly interesting research topic of its own, presently
pursued by quite a few people (see the papers by Rieffel or Weaver
\cite{Rieffel1},\cite{Weaver} and the references mentioned therein).
Another, earlier (and important) source is \cite{Davies}.  We recently
extended the investigation of metric structures to socalled \tit{lump
  spaces} and \tit{probabilistic metric spaces} (see \cite{requroy}
and \cite{schweizer}) and employed it in the general context of
quantum gravity (cf. also \cite{2}).\\[0.3cm]
Remark: When we wrote an earlier draft of this paper we were unaware
of the content of \cite{Davies}. It happened only recently that we
realized that various of the results we derived in connection with
Dirac-operators and the Connes-distance can already be found in
\cite{Davies} and we try to take care of this fact in the following.
Both the technical approach and the motivation are, however, not
entirely identical. The interested reader may consult the electronic
version of our earlier draft (\cite{graph}).

The same applies to paper \cite{Mohar} where some of the Hilbert space
methods were developed we later rederived in \cite{graph} being
unaware of the prior results. As a consequence we drop most of the
technical steps leading to that part of our previous results
overlapping with corresponding parts in the above mentioned papers and
refer, for simplicity to \cite{graph} where the interested reader can
find more details.

\section{A brief Survey of Differential and Operator\\ Calculus on
  Graphs and Graph\\ Hilbert Spaces}

We give a brief survey of certain concepts and tools needed
in the following analysis. While our framework may deviate at various
places from the more traditional one, employed in e.g.
\tit{algebraic graph theory} (see e.g. \cite{8},\cite{9} or
\cite{Colin}), this is mainly done for reasons of greater mathematical
flexibility and generality and, on the other side, possible physical
applications (a case in point being the analysis of
\tit{non-commutative spaces}). Most of the technical tools which are
not defined in detail in the following have been introduced in section 3 of
\cite{1}.
\subsection{Simple (Symmetric) Graphs}
For simplicity we assume the graph to be connected and \tit{locally
  finite} (no elementary loops and multiedges, whereas these could
easily be incorporated in the framework), i.e. each node (or vertex)
is incident with only a finite number of edges (or bonds). To avoid
operator domain problems we usually make the even stronger assumption
that the \tit{vertex--} or \tit{node degree}, $v_i$, ($i$ labelling the nodes), is
globally bounded but this restriction is frequently not really
necessary. Furthermore, it has turned out to be algebraically
advantageous to identify an (undirected) labelled graph with a
directed graph having two oppositely directed edges for each
undirected edge, the directed edge, pointing from node $n_i$ to node
$n_k$, being denoted by $d_{ik}$, the oppositely directed edge by
$d_{ki}$ and the undirected (but orientable) edge by their
superposition
\begin{equation}b_{ik}=d_{ik}-d_{ki}=-b_{ki}\end{equation}
($d_{ik}$ and $d_{ki}$ are treated as independent basis vectors;
cf. \cite{1} or \cite{graph}).

As the elementary building blocks of our graph Hilbert spaces we take
$\{n_i\}$ and $\{d_{ik}\}$ as basis elements of a certain hierarchy of
Hilbert spaces over, say, $\C$ with scalar product induced by
\begin{equation}
(n_i|n_k)=\delta_{ik}\quad(d_{ik}|d_{lm})=\delta_{il}\cdot\delta_{km}\end{equation}
This definition implies $(b_{ik}|b_{ik})=2$.
\begin{defi}[Vertex-, Edge Hilbert Space]
  The Hilbert spaces $H_0$, $H^a_1$ ($a$ for
  antisymmetric) and $H_1$ consist of the formal sums
\begin{equation} f:=\sum f_in_i\quad g:=\sum
  g_{ik}d_{ik}\quad\mbox{with}\quad g_{ik}=-g_{ki}\quad\mbox{and}\quad
 g':=\sum g_{ik}d_{ik} \end{equation}
\begin{equation} \sum |f_i|^2<\infty\quad \sum|g_{ik}|^2<\infty\end{equation}
$f_i,g_{ik}$ ranging over a certain given field like e.g. $\C$
(sometimes only rings like e.g. $\Z$ are admitted; then we are dealing
only with modules). We evidently have $H^a_1\subset H_1$.
\end{defi}
Remark: One could continue this row of vector spaces in ways which are
common practice in, say, {\em algebraic topology} ( see \cite{1}
sections 3.1 and 3.2).  In this context they are frequently called
{\em chain complexes} (see also \cite{Novikov}).  On the other hand,
the above vector spaces could as well be viewed as {\em discrete
  function spaces} over the {\em node-, bond set} with $n_i,d_{ik}$
now representing the elementary {\em indicator functions}.
\vspace{0.3cm}

Proceding in this spirit we can now introduce two linear maps between
$H_0,H_1$ extending the usual \tit{boundary-} and
\tit{coboundary map}. On the basis elements they act as follows:
\begin{equation} \delta:\; d_{ik}\to n_k\quad\text{hence}\quad b_{ik}\to
n_k-n_i\end{equation}
\begin{equation} d:\; n_i\to \sum_k(d_{ki}-d_{ik})=\sum_k b_{ki}\end{equation}
and linearly extended. That is, $\delta$ maps the directed bonds
$d_{ik}$ onto the terminal node  and $b_{ik}$ onto its (oriented) {\em boundary},
while $d$ maps the node $n_i$ onto the sum of the {\em ingoing} directed
bonds minus the sum of the {\em outgoing} directed bonds or on the sum of
{\em oriented} {\em ingoing} bonds $b_{ki}$.

As was shown in \cite{1} (we later realized that the same definition
was already employed in \cite{Davies}), these definitions lead in fact to a
kind of \tit{discrete differential calculus} on $H_0,H_1$, that is we
have
\begin{equation} df=d(\sum f_in_i)=\sum_{k,i}(f_k-f_i)d_{ik}\end{equation}

Combining now the operators $\delta$ and $d$, we can construct the \tit{canonical graph Laplacian}. On the vertex space it reads:
\begin{equation} \delta df=-\sum_i(\sum_k f_k-v_i\cdot
  f_i)n_i=-\sum_i(\sum_k(f_k-f_i))n_i=:-\Delta f \end{equation} where
$v_i$ denotes the {\em node degree} or {\em valency} defined above and
the $k$-sum extends over the nodes adjacent to $n_i$.\\[0.3cm]
Remark: Note that there exist several variants in the literature (see e.g.
\cite{Colin} or \cite{9}). Furthermore, many mathematicians
employ a different sign-convention. We stick in the following to the
convention being in use in the mathematical-physics literature
where $-\Delta$ is the positive(!) operator.
\vspace{0.3cm}

This \tit{graph Laplacian} is intimately connected with another
important object, employed in algebraic graph theory, i.e. the {\it
  adjacency matrix}, $A$, of a graph, its entries, $a_{ik}$, having the
value one if the nodes $n_i,n_k$ are connected by a bond and are zero
elsewhere. If the graph is {\em undirected} (but orientable), the relation between $n_i,n_k$ is {\em symmetric}, i.e.
\begin{equation} a_{ik}=1\quad\Rightarrow\quad a_{ki}=1\end{equation}
This has the obvious consequence that in case the graph is {\em
  simple} and {\em undirected}, $A$ is a
symmetric matrix with zero diagonal elements.
\\[0.3cm]
Remark: More general $A$'s occur if more general graphs are
admitted (e.g. general multigraphs).
\\[0.3cm]
With our definition of $\Delta$ it holds:
\begin{equation} \Delta=A-V\end{equation}
where $V$ is the diagonal {\em degree matrix}, having $v_i$ as
diagonal entries.
(Note that the other sign-convention would lead to $\Delta=V-A$).

To avoid domain problems we assume from now on that the {\it node
  degree}, $v_i$, is {\it uniformly bounded} on the graph $G$, i.e.
\begin{equation} v_i\le v_{max}<\infty\end{equation}
Defining $d_{1,2}$ as
\begin{equation} d_{1,2}:\;n_i\to\sum d_{ki}\; ,\;\sum d_{ik}\end{equation}
respectively and linearly extended, we get
\begin{equation} d=d_1-d_2\end{equation}
Similarly we make the
identification $\delta=:\delta_1$ with
\begin{equation} \delta_{1,2}:\;d_{ik}\to n_k,\,n_i\end{equation}

It is noteworthy (but actually not surprising) that $v_i\le v_{max}$
implies that all the above operators are {\it bounded} (in contrast to
their continuous counterparts, which are typically unbounded). Taking
this for granted at the moment, a straightforward analysis yields the
following relations:
\begin{lemma}\hfill \label{newrelations}
\begin{enumerate}
\item The adjoint $d^{\ast}$ of $d$ with respect to
the spaces $H_0,H^a_1$ is $2\delta$
\item On the other side we have for the natural extension of
  $d,\delta$ to the larger space
$H_1$:
\begin{equation} \delta_1=(d_1)^{\ast}\;,\;\delta_2=(d_2)^{\ast}\end{equation}
hence
\begin{equation} (\delta_1-\delta_2)=(d_1-d_2)^{\ast}=d^{\ast}\neq 2\delta=2\delta_1\end{equation}
\item Furthermore it holds
\begin{equation}d_1^*\cdot d_1=\delta_1\cdot d_1=d_2^*\cdot
  d_2=V:\;n_i\to v_in_i\end{equation}
(with $V$ the vertex degree matrix)
\begin{equation}d_1^*\cdot d_2=\delta_1\cdot d_2=d_2^*\cdot
  d_1=\delta_2\cdot d_1=A:\; n_i\to\sum_{k-i}n_k\end{equation}
Similar geometric properties of the graph are encoded in the products
coming in reversed order.
\end{enumerate}
\end{lemma}
(Here and in the following $k-i$ means summation over the first index
and runs through the set of labels of nodes directly connected with $n_i$).

That and how $d,d^*$ encode certain geometric information about the
graph can be seen from the following domain- and range-properties (cf.
\cite{graph}, for corresponding results in the more traditional
approach see \cite{8},p.24ff).
\begin{satz}Let the graph be connected and finite, $|\mcal V|=n$, then
\begin{equation}dim(Rg(d^*))=n-1\end{equation}
\begin{equation}dim(Ker(d^*))=\sum_i v_i-(n-1)\end{equation}
With $dim(H_1)=\sum_i v_i$, $dim(H_1^a)=1/2\cdot dim(H_1)$ we have
\begin{equation}codim(Ker(d^*))=dim(Rg(d))=n-1\end{equation}
We see that both $Rg(d^*)$ and $Rg(d)$ have the same dimension $(n-1)$.
\end{satz}
\begin{bem}In case the graph has, say, $c$ components, the above
  results are altered in an obvious way; we have for example
\begin{equation}dim(Rg(d^*))=n-c\end{equation}
\end{bem}
In the literature $Ker(d^*)$ is called (for obvious reasons)
  the {\em cycle subspace} (cf e.g. \cite{8}). On the antisymmetric
  subspace $H_1^a$ we have $d^*=2\delta$ and
  $\delta(b_{ik})=n_k-n_i$. Choosing now a {\em cycle}, given by its
  sequence of consecutive vertices $n_{i_1},\ldots,
  n_{i_k};n_{i_{k+1}}:=n_{i_1}$, we have
\begin{equation}d^*(\sum
  b_{i_li_{l+1}})=2\sum(n_{i_{l+1}}-n_{i_l})=0\end{equation}
that is, vectors of this kind lie in the kernel of $d^*$

We will now provide quantitative lower and upper bounds for the respective
norms of the occurring operators. We have:
\begin{multline}
  \|df\|^2=\sum_{ik}|(f_k-f_i)|^2=\sum_iv_i\cdot|f_i|^2+\sum_kv_k\cdot|f_k|^2-\sum_{i\neq
    k}(\overline{f_k}f_i+\overline{f_i}f_k)\\
=2\cdot\sum_i
  v_i|f_i|^2-2\cdot\sum_{i\neq k}\overline{f_k}f_i
\end{multline}
which can be written as:
\begin{equation} \|df\|^2=2((f|Vf)-(f|Af))=(f|-2\Delta f)\end{equation}
and shows the close relationship of the norm of $d$ with the {\it
  expectation values} of the {\it adjacency} and {\it degree matrix}
or the {\it graph Laplacian}. That is, norm estimates for,
say, $d$, derive in a natural manner from the corresponding estimates
for $A$ or $-\Delta$. With
\begin{equation} \|df\|^2=(f|d^{\ast}df)=(f|-2\Delta f)\end{equation}
we have
\begin{equation}
0\leq d^{\ast}d\,=\,-2\Delta\quad\text{and}\quad\|d\|^2=\sup_{\|f\|=1}(f|-2\Delta
f)=\|-2\Delta\|\end{equation}
Furthermore via
\begin{equation} 0<\sup_{\|f\|=1}(f|-2\Delta f)\leq
2v_{max}+2\sup_{\|f\|=1}|<f|Af>|\end{equation}
we get
\begin{equation}\|-\Delta\|\leq v_{max}+\|A\| \end{equation}
Remark: We want to mention that we are  using the usual {\em operator norm}
also for matrices (in contrast to most of the matrix literature),
which is also called the {\it spectral norm}. It is unique in so far
as it coincides with the so-called {\it spectral radius} (cf. e.g.
\cite{11} or \cite{12}), that is
\begin{equation} \|A\|:=\sup\{|\lambda|;\,\lambda\in spectr(A)\}\end{equation}

In a first step we give upper and lower bounds for the operator norm of the
adjacency matrix, $A$, both in the finite- and infinite-dimensional
case. There are various proofs available of a varying degree of
generality (see e.g. \cite{Mohar}, \cite{Colin} or \cite{graph}) to
which we refer the reader. In the following we give only the final
results. Note however that the transition from finite to infinite
graphs is far from straightforward as in some of the necessary
technical steps entirely new methods are needed.
\begin{satz}[Norm of $A$] With the adjacency matrix $A$
finite or infinite and a finite $v_{max}$ we have the following
result (a certain fixed labelling of the nodes being assumed):
\begin{equation}\limsup\, n^{-1}\cdot\sum_{i=1}^n
  v^{(n)}_i\leq\limsup\|A_n\|=
\|A\|=\sup\{|\lambda|;\,\lambda\in\;spectr(A)\}
\leq v_{max}\end{equation}
\end{satz}
Here $A_n$ are the adjacency matrices for the induced subgraphs,
living over the first $n$ labelled nodes, $v_i^{(n)}$ is the
corresponding induced (and $n$-dependent) node degree.

As a byproduct we have the important lemma
\begin{lemma}The adjacency matrices, $A_n$, converge strongly to $A$
  and we have in particular $\|A_n\|\nearrow\|A\|$.
\end{lemma}
(For a proof of the latter result see \cite{Mohar} or \cite{graph})
\begin{bem}To prove strong convergence of operators is of some
  relevance for the limit behavior of {\em spectral properties} of the
  operators $A_n,A$. That is (cf. e.g. \cite{Reed} section VIII.7), we
  have in that case ($A_n,A$ selfadjoint and uniformly bounded)
  $A_n\to A$ in {\em strong resolvent sense}, which implies that the
  spectrum of the limit operator, $A$, cannot suddenly expand, i.e.
\begin{equation}\lambda\in spec(A)\Rightarrow\exists\; \lambda_n\in
  spec(A_n)\; \text{with}\; \lambda_n\to \lambda\end{equation}
and for $a,b\not\in spec_{pp}(A)$
\begin{equation}P_{(a,b)}(A_n)\to
  P_{(a,b)}(A)\;\text{strongly}\end{equation}
\end{bem}

To test the effectiveness of the upper and lower bounds given above,
we apply them to a non-trivial model recently discussed in
\cite{Froese}, i.e. the infinite \tit{binary tree} with \tit{root}
$n_0$ where $v_0$ is two and $v_i$ equals three for $i\neq 0$. The
authors show (among other things) that the spectrum consists of the
interval $[-2\sqrt{2},2\sqrt{2}]$, i.e. $\|A\|=2\sqrt{2}$. $v_{max}$
is three, we have to calculate $\limsup 1/n\cdot\sum_1^n v_i$. For
simplicity we choose a subsequence so that $n:=n(N)$ with $N$ denoting
the $N$-th level (consisting of $2^N$ nodes) of the tree starting from
the root $n_0$. Note that in the corresponding \tit{induced subgraph}
$G_N$ the \tit{boundary nodes} sitting in the $N$-th level have only
\tit{node degree} one with respect to $G_N$ but three viewed as nodes
in the full tree.

We then have
\begin{equation}n=\sum_{k=1}^N 2^k\; , \; \sum_{i=0}^{n(N)}
  v_i=2+3\cdot\sum_{k=1}^{N-1}2^k+2^N=3\cdot\sum_{k=0}^N 2^k
-2\cdot2^N - 1\end{equation}
Hence
\begin{equation}\lim_{n(N)} 1/n(N)\sum_{i=0}^{n(N)}=3-2\lim_N
  (\sum_0^N 2^{k-N})^{-1}=2\end{equation}
That is, our general estimate imply $2\leq\|A\|\leq3$, which is not so bad.
\subsection{Arbitrary Directed Graphs}
If we deal with general directed graphs we have both ingoing and
outgoing edges at each node but in general they do no longer occur in
a symmetric way. But nevertheless, most of our concepts
and tools, developed in the foregoing subsection, do still exist. The
definitions of $d$ and $d^*$ are unaltered. As each edge, $d_{ik}$, is
an \tit{outgoing} edge for node $n_i$ and an \tit{ingoing} edge for
node $n_k$, the same expression holds for $df$, i.e.
\begin{equation}df=\sum_{i,k} (f_i-f_k)d_{ik}\quad\text{with}\quad
  f=\sum_i f_in_i\end{equation}
(the sum of course only extends over those directed edges which do
exist in the directed graph; in particular each edge, $d_{ik}$, is
only counted once in the sum as an outgoing edge with respect to the
label $i$.)
Furthermore the notions of $d_{1,2}$ and $\delta_{1,2}$ remain the same, mapping
nodes or
edges on ingoing edges, outgoing nodes and vice versa. We again have
\begin{equation}\delta_{1,2}=d_{1,2}^*\quad\text{and}\quad (\delta_1-\delta_2)=d^*=(d_1-d_2)\end{equation}

We can now calculate $d_1^*\cdot d_1$ and $d_2^*\cdot d_2$ and get:
\begin{equation}d_1^*\cdot d_1(n_i)=v_i^{in}\cdot n_i\quad ,\quad d_2^*\cdot
  d_2(n_i)=v_i^{out}\cdot n_i\end{equation}
with $v_i^{in,out}$ the \tit{in-,out-degree} of the node $n_i$.
In the same way we can calculate
\begin{equation}\delta_1\cdot d_2(n_i)=\sum_{k-i,out}n_k\quad ,\quad
  \delta_2\cdot d_1(n_i)=\sum_{k-i,in}n_k\end{equation}
This yields
\begin{equation}d^*d=(\delta_1-\delta_2)\cdot(d_1-d_2)=(V^{in}+V^{out})-(A^{in}+A^{out})\end{equation}
where the occurring operators on the rhs are the \tit{in-,out-vertex degree
  matrices}, \tit{in-,out-adjacency matrices} respectively.

In general the individual in-,out-adjacency matrices are
non-symmetric. Their sum, however, is symmetric (and is, in the case
of a symmetric graph, twice the adjacency
matrix of the undirected graph, i.e. $2A$). One can now
again define a (positive, i.e. symmetric) Laplace operator for a non-symmetric (directed) graph,
that is:
\begin{conclusion}
\begin{equation}-\Delta:=d^*d=(V^{in}+V^{out})-(A^{in}+A^{out})=:V_d-A_d\end{equation}
\end{conclusion}
(Note the now missing factor two in front of $\Delta$!)
\section {\label{Triplet}The Spectral Triplet on a general
  (directed or undirected) Graph} We begin by making some remarks on
various concepts, being in use in the more recent literature. We note
that our version of a Dirac operator (defined below) intertwines
node-vectors and bond-vectors while in other examples it maps node- to
node-functions. Our bond-functions have (in some sense) the character
of \tit{cotangential-vectors}, while in other approaches derivatives
of functions are interpreted as \tit{tangent-vectors}. In our view,
the latter formalism is only effective in certain classes of highly
regular models (like e.g. lattices) where one has kind of global
directions and will become cumbersome for general graphs. We developed
this latter approach a little bit in section 3.3 of \cite{1} and
showed how these cotangent and tangent vectors can be mapped into each
other. We think that, on the other side, our framework is more
flexible in the general case. This holds in particular for our Dirac
operator, which encodes certain geometric properties of the underlying
discrete ``manifold''.

The Hilbert space we will use in the following is
\begin{equation} H=H_0\oplus H_1\end{equation}
The {\it natural representation} of the function algebra ${\cal F}$
(consisting of the bounded node functions)
\begin{equation} \{f;f\in {\cal C}^\infty\;\text{i.e.}\;\sup_i |f_i|<\infty\}\end{equation}
on $H$ by bounded operators is given by:
\begin{equation} H_0:\;f\cdot f'=\sum f_if'_i\cdot n_i\;\mbox{for}\;f'\in H_0
\end{equation}
\begin{equation} H_1:\;f\cdot\sum g_{ik}d_{ik}:=\sum f_ig_{ik}d_{ik}\end{equation}
From previous work (\cite{1}) we know that $H_1$ carries also
a right-module structure, given by:
\begin{equation} \sum g_{ik}d_{ik}\cdot f:=\sum g_{ik}f_k\cdot d_{ik}\end{equation}
(For convenience we do not distinguish notationally between
elements of $\Fc$ and their Hilbert space
representations).\\[0.3cm]
Remark: The same bi-module structure and the Dirac operator defined
below were already employed in
\cite{Davies}, p.414ff.
\vspace{0.3cm}

An important object in various areas of modern analysis on manifolds
or in Connes' approach to noncommutative geometry is the so-called {\it Dirac
  operator} $D$ (or rather, a certain version or variant of its
classical counterpart; for the wider context see
e.g. \cite{4} or \cite{Landi} to \cite{Esposito}). As $D$ we will take
in our context the operator:
\begin{equation} D:=\left( \begin{array}{cc}0 & d^{\ast}\\d & 0 \end{array} \right)\end{equation}
acting on
\begin{equation} H=\left( \begin{array}{c}H_0 \\ H_1 \end{array} \right)\end{equation}
with
\begin{equation} d^{\ast}=(\delta_1-\delta_2)\end{equation}
Note however, that there may exist in general several
possibilities to choose such an operator. On the other hand, we
consider our personal choice to be very natural from a geometrical
point of view.
\begin{lemma}There exists in our scheme a natural chirality- or
  grading operator, $\chi$ and an antilinear involution, $J$. given by
\begin{equation}\chi:=\left( \begin{array}{cc}1 & 0\\0 & -1 \end{array} \right)\end{equation}
with
\begin{equation}[\chi,{\Fc}]= 0    \quad \chi\cdot
  D+D\cdot\chi= 0\end{equation}
and
\begin{equation}J:\left( \begin{array}{c}x \\ y \end{array} \right)\to\left( \begin{array}{c}\overline{x} \\\overline{y} \end{array} \right)
\end{equation}
so that
\begin{equation}J\cdot f\cdot J=\overline{f}\end{equation}
\end{lemma}
These are some of the ingredients which establish what
Connes calls a \tit{spectral triple} (cf. e.g. \cite{Grav} or
\cite{Finite}). We do not want, however, to introduce the full
machinery at the moment as our scheme has an independent geometric
meaning of its own. Note in particular what we are saying below about
(non) compactness of various operators in observation \ref{compact}.
\begin{defi}[Spectral Triplets] As {\em spectral
  triplet} on a general graph we take
\begin{equation} (H,{\Fc},D)\end{equation}
\end{defi}

In our general framework we got in a relatively straightforward manner
a Hilbert space being the direct sum of the \tit{node space} (a
function space) and the \tit{bond space} (resembling the set of
cotangent vectors) and a Dirac operator which emerged naturally as
kind of a square root of the Laplacian.

On the other side, if one studies simple models as e.g. in \cite{5} to
\cite{7}, other choices are possible. In \cite{5},\cite{6}, where the
one-dimensional lattice was studied, the \tit{symmetric difference
  operator} was taken as Dirac operator. In \cite{7} the
one-dimensional lattice was assumed to be directed (i.e., in our
notation, only $d_{i,i+1}$ were present) and the Dirac operator was
defined (somewhat adhoc) as a certain self adjoint ``doubling'' of the
(one-sided, i.e.  non-symmetric) adjacency matrix. As we will show
below, this latter model fits naturally in our general approach which
includes both directed and undirected graphs. All these Dirac
operators are different and it is hence no wonder that they lead to
different consequences (see below). It is our opinion that, in the
end, an appropriate choice has to be dictated by physical intuition.
Nevertheless, this apparent non-uniqueness should be studied more
carefully.

As can be seen from the above, the connection with the graph Laplacian
is relatively close since for e.g. a symmetric graph we have:
\begin{equation} D^2=\left( \begin{array}{cc}\dast d & 0\\0 & d\dast \end{array}
\right)\end{equation}
and
\begin{equation} \dast d=-2\Delta\end{equation}
$d\dast$ is the corresponding object on $H_1$. (In the vector analysis
of the continuum the
two entries correspond to $\operatorname{divgrad}\,
,\,\operatorname{graddiv}$ respectively ).

In the original approach of Connes great emphasis was laid on the
compactness of operators like the inverse of the Dirac operator and it
is part of the general definition of a \tit{spectral triple}. As
discrete spaces of the kind we are studying are non-trivial examples
of non-commutative spaces, it is interesting that we can easily test
whether this assumption is fulfilled in our particular setting. As may
be expected, for graphs with globally bounded node degree, we have the
following result:
\begin{ob}\label{compact}Note that all our operators are bounded, the Hilbert space
  is (in general) infinite dimensional, hence there is no chance to
  have e.g. $(D-z)^{-1}$ or $(D^2-z)^{-1}$ compact. At the moment we
  are sceptical whether this latter phenomenon dissappears generically
  if the vertex degree is allowed to become infinite.  There are some
  results on spectra of random graphs which seem to have a certain
  bearing on this problem (cf. e.g. \cite{Juhasz}).
\end{ob}

In the next sections we introduce and calculate the socalled
Connes-distance functional and compare it, among other things,
with the ordinary graph distance. In doing this we have to calculate
the commutator $[D,f]$ applied to an element $f'\in H_0$. We have:
\begin{equation} (d\cdot f)f'=\sum_{ik}(f_kf'_k-f_if'_i)d_{ik}\end{equation}
\begin{equation} (f\cdot d)f'=\sum_{ik}f_i(f'_k-f'_i)d_{ik}\end{equation}
hence
\begin{equation} [D,f]f'=\sum_{ik}(f_k-f_i)f'_kd_{ik}\end{equation}
On the other side the right-module structure allows us to define $df$
as an operator on $H_0$ via:
\begin{equation} df\cdot f'=(\sum_{ik}(f_k-f_i)d_{ik})\cdot (\sum_k f'_kn_k)=\sum_{ik}(f_k-f_i)f'_kd_{ik}=[D,f]f'\end{equation}
In a next step we define $df$ as operator on $H_1$ which is not as
natural as on $H_0$. We define:
\begin{equation} df|_{H_1}:\;d_{ik}\to (f_i-f_k)n_k\end{equation}
and linearly extended. A short calculation shows
\begin{equation}df|_{H_1}=-(d\bar{f}|_{H_0})^*=[d^*,f] \end{equation}
with
\begin{equation}[d^*,f]g=d^*(f\cdot g)-fd^*g\quad (g\in H_1)\end{equation}
This then has the following desirable consequence:
\begin{conclusion} With the above definitions the representation of $df$ on
  $H$ is given by
\begin{equation}df|_H=\begin{pmatrix}0 & df|_{H_1} \\ df|_{H_0} &
    0\end{pmatrix}=\begin{pmatrix}0 & -(d\bar{f}|_{H_0})^* \\
    df|_{H_0} & 0\end{pmatrix}  \end{equation}
and it immediately follows
\begin{equation}df|_H=\begin{pmatrix}0 & [d^*,f] \\ {[}d,f] & 0\end{pmatrix}=  [D,f] \end{equation}
\end{conclusion}

\section{The Connes-Distance Function on Graphs}
From the general theory of operators on Hilbert spaces we know that:
\begin{equation}\label{eqqq} \|T\|=\|T^{\ast}\|\end{equation}
Hence
\begin{lemma}
\begin{equation}\label{eq} \|[d,f]\|= \|[d,\bar{f}]\|=\|[d^*,f]\|\end{equation}
and
\begin{equation} \|[D,f]\|=\|[d,f]\|\end{equation}
\end{lemma}
Proof: The left part of (\ref{eq}) is shown below and is a consequence
of formula (\ref{eqq}); the right identity follows from (\ref{eqqq}). With
\begin{equation} X:=\left( \begin{array}{c}x \\ y \end{array} \right)\end{equation}
and $T_1:=[d,f]$,\;$T_2:=[d^*,f]$, the norm of $[D,f]$ is:
\begin{equation}
\|[D,f]\|^2=\sup\{\|T_1x\|^2+\|T_2y\|^2;\,\|x\|^2+\|y\|^2=1\}\end{equation}
Normalizing now $x,y$ to $\|x\|=\|y\|=1$ and representing a general
normalized vector $X$ as:
\begin{equation} X=\lambda x+\mu
y\;,\;\lambda,\mu>0\;\mbox{and}\;\lambda^2+\mu^2=1 \end{equation}
we get:
\begin{equation}
\|[D,f]\|^2=\sup\{\lambda^2\|T_1x\|^2+\mu^2\|T_2y\|^2;\|x\|=\|y\|=1,\lambda^2+\mu^2=1\}\end{equation}
where now $x,y$ can be varied independently of $\lambda,\mu$ in
their respective admissible sets, hence:

\begin{equation}
  \|[D,f]\|^2=\sup\{\lambda^2\|T_1\|^2+\mu^2\|T_2\|^2\}=\|T_1\|^2 \quad
\Box \end{equation}
(as a consequence of equation (\ref{eq})).

We see that in calculating $\|[D,f]\|$ we can restrict ourselves to
the simpler expression $\|[d,f]\|$. We infer from the above
calculations ($x\in H_0$):
\begin{equation}\label{eqq} \|df\cdot x\|^2=\sum_i(\sum_{j=1}^{v_i}|f_i-f_{k_j}|^2\cdot|x_i|^2\end{equation}
and the corresponding expression for a directed graph with $v_i$
replaced by $v_i^{out}$.
Abbreviating
\begin{equation} \sum_{j=1}^{v_i}|f_{k_j}-f_i|^2=:a_i\geq 0\end{equation}
and calling the supremum over $i$ $a_s$, it follows:
\begin{equation} \|df\cdot x\|^2= a_s\cdot(\sum_i a_i/a_s\cdot|x_i|^2)\leq
a_s\end{equation}
for $\|x\|^2=\sum_i |x_i|^2=1$.

On the other side, choosing an appropriate sequence of normalized basis vectors $e_{\nu}$ so
that the corresponding $a_{\nu}$ converge to $a_s$ we get:
\begin{equation} \|df\cdot e_{\nu}\|^2\to a_s\end{equation}
We hence have
\begin{satz} \label{norm}
\begin{equation}
  \|[D,f]\|=\sup_i(\sum_{j=1}^{v_i}|f_{k_j}-f_i|^2)^{1/2}\; ,\; \sup_i(\sum_{j=1}^{v_i^{out}}|f_{k_j}-f_i|^2)^{1/2}\quad\text{respectively}\end{equation}
\end{satz}

The \tit{Connes-distance functional} between two nodes, $n,n'$, is now defined as
follows:
\begin{defi}[Connes-distance function]
\begin{equation} dist_C (n,n'):=\sup\{|f_{n'}-f_n|;\|[D,f]\|=\|df\|\leq 1\}\end{equation}
\end{defi}
We would like to note that Davies in \cite{Davies} introduced several
metrics on graphs, which have been motivated, as he remarks, by his
study of heat kernels on Riemannian manifolds. What he calls metric
$d_3$, is related to the rhs of the equation in theorem \ref{norm} (he
uses slightly different Hilbert spaces). He then shows in a longer
proof that this metric is identical to another one, $d_4$, which
corresponds to the lhs in the above theorem. In our approach, on the
other side, the content of theorem \ref{norm} is derived in a
relatively transparent and straightforward way.
\begin{bem} It is easy to prove that this defines a metric on the
graph.
\end{bem}
\begin{koro} It is sufficient to vary only over the set
$\{f;\|df\|=1\}$.
\end{koro}
Proof: This follows from
\begin{equation} |f_k-f_i|=c\cdot|f_k/c-f_i/c|\;;\;c=\|df\|\end{equation}
and
\begin{equation} \|d(f/c)\|=c^{-1}\|df\|=1\end{equation}
with $c\leq1$ in our case.\bewende
\vspace{0.5cm}

In general it turns out to be a nontrivial task to calculate this
distance on an arbitrary graph as the nature of the above constraint
is quite subtle . The underlying reason is that the constraint is, in
some sense, inherently \tit{non-local}. As $f$ is a function, the
difference, $f_{n'}-f_n$, has to be the same independently of the path
we follow, connecting $n'$ and $n$. On the other side, in a typical
optimization process one usually deals with the individual jumps,
$f_k-f_i$, between neighboring points along some path. It is then not
at all clear that these special choices of jumps along such a path can
be extended to a global function without violating the overall
constraint on the expression in theorem \ref{norm}.  Nevertheless we
think the above closed form is a solid starting point for the
calculation of $dist_C$ on various classes of graphs or lattices. We
illustrate this by proving some apriori estimates concerning this
distance function and by evaluating it for some examples.
\subsection{Some General Estimates}
Having an admissible function $f$ so that
$\sup_i(\sum_{k=1}^{v_i}|f_k-f_i|^2)^{1/2}\leq 1$, this implies that,
taking a {\it minimal path} $\gamma$ from, say, $n$ to $n'$, the
jumps $|f_{\nu+1}-f_{\nu}|$ between neighboring nodes along the path
have to fulfill:
\begin{equation} |f_{\nu+1}-f_{\nu}|\leq 1\end{equation}
and, a fortiori, have to be strictly smaller than $1$ in the general
situation.

On the other side the Connes distance can only become
identical to the ordinary distance $d(n,n')$ if there exist a sequence
of admissible node functions with all these jumps approaching the
value $1$ along such a path, which is however impossible in general as
can be seen from the structure of the constraint on the expression in
theorem \ref{norm} . Only in this case one gets:
\begin{equation}| \sum_{\gamma}(f_{\nu+1}-f_{\nu})|\to
\sum_{\gamma}1=length(\gamma)\end{equation}
We formulate this observation as follows:
\begin{lemma}[Connes-distance]Within our general scheme one has the following
  inequality
\begin{equation} dist_C(n,n')\leq d(n,n')\end{equation}
By the same token one can prove that $dist_C$ between two nodes is
bounded by the corresponding Connes-distance calculated for the
(one-dimensional) sub-graph formed by a minimal path connecting these
nodes, i.e.
\begin{equation}dist_C(n,n')\leq dist_C(min.path)(n,n')\end{equation}
\end{lemma}
The reason is that one has more \tit{admissible functions} at ones
disposal for a subgraph. With $G'$ a connected subgraph of $G$, the
set of admissible function, $S_{G'}$, on $G'$ contains the
restrictions of the functions of the corresponding set, $S_G$,
belonging to $G$, as each restriction to $G'$ of a member belonging to
$S_G$ lies in $S_{G'}$.  Hence the supremum is in general larger on
$S_{G'}$. The distance along such a path, on the other side, can be
rigorously calculated (see the discussion of some examples below) and
is for non-neighboring nodes markedly smaller than the ordinary graph
distance. From what we have said we can also infer the following
corollary.
\begin{koro}With $G'$ a connected
  subgraph of $G$ it holds (with $n,n'\in V'\subset V$)
\begin{equation}dist_C(n,n';G)\leq dist_C(n,n';G')\end{equation}
\end{koro}

One can also give sufficient criteria for $dist_C(n,n')<d(n,n')$. The
cases of undirected, directed graphs, respectively, have to be treated
a little bit differently.
\begin{lemma}\label{equal}Let $G$ be an undirected graph and $\gamma$ a minimal
  path of length, $l>1$, connecting $n,n'$. There is at least one
  node, $n^*$, belonging to $\gamma$, having node degree $\geq2$ (as
  there are at least two consecutive edges, belonging to $\gamma$). If
\begin{equation}dist_C(n,n')=l=d(n,n')\end{equation}
all the individual jumps along $\gamma$ have to be one. But then the
corresponding function cannot be admissible at $n^*$. Hence
\begin{equation}dist_C(n,n')<d(n,n')\end{equation}

Let $G$ now be a directed graph and let there exist two different
paths, $\gamma,\gamma'$, of equal length, $l>1$, connecting
$n,n'$. Again there exists a node, $n^*$, on $\gamma$ so that it is
incident with two edges, the one belonging to $\gamma$, the other to
$\gamma'$. Along both edges the jumps have to be one and again the
admissibility of the corresponding function is violated. We again can
conclude
\begin{equation}dist_C(n,n')<d(n,n')\end{equation}
\end{lemma}
Remark:The latter situation will be discussed below in the example of
the directed $\Z^2$-lattice.\vspace{0.3cm}

We remarked above that the calculation of the Connes distance on
graphs is to a large part a continuation problem for admissible
functions, defined on subgraphs. Then the following question poses itself.
For what classes of graphs and/or subgraphs do we have an equality in
the above corollary? We start from a given graph, $G_0=(V_0,E_0)$, and then
add new nodes and bonds, yielding a new graph, $G'=(V',E')$. We
consider two fixed nodes, $n_0,n_0'$ in $G_0$.
\begin{assumption} We assume that the above process does not create
  new paths between nodes belonging to $G_0$. In other words, the paths,
  connecting $n_0$ and $n_0'$ are contained in $G_0$.
\end{assumption}
\begin{lemma}Under this assumption each admissible function on $G_0$
  can be extended to an admissible function on $G'$.
\end{lemma}
Proof: In a first step we construct the set of \tit{nearest
  neighbors}, $V_1\backslash V_0$ in $V'\backslash V_0$ relative to
$V_0$. Each new node in $V_1\backslash V_0$ has a unique nearest
neighbor in $V_0$ since otherwise there would exist a new path between
these two nodes lying in $G_0$. With $n\in V_1\setminus V_0$ we extend an
admissible function on $G_0$ as follows:
\begin{equation}f_n:=f_{n_0}\quad\text{$n_0$ the unique nearest
    neighbor in $V_0$}\end{equation}
This extended function is an admissible function on
$G_1:=(V_1,E_1)$. Note however that, by assumption, there do not exist
bonds in $E_1$, connecting nodes in  $V_1\backslash V_0$. We can now
continue this process until we arrive at the graph $G'$.\bewende
\\[0.3cm]
By the same token we see that
\begin{equation}dist_C(n_0,n_0';G_1)=dist_C(n_0,n_0';G_0)\end{equation}
This holds at every intermediate step and we get:
\begin{lemma}Under the above assumption we have
\begin{equation}dist_C(n_0,n_0';G_0)=dist_C(n_0,n_0';G')\end{equation}
\end{lemma}
\begin{koro}If $G$ is a tree, it holds
\begin{equation}dist_C(n;n')=dist_C(n;n';\text{minimal
    path})\end{equation}
\end{koro}
Proof: In a tree there exists, by definition, at most one path,
connecting two nodes. We can take this path as connected subgraph,
$G_0$, and make the above extension, since $G$ and $G_0$ fulfill the
assumption.\bewende\\[0.3cm]
The graphs, so constructed are however rather special, consisting, so
to speak, of a start graph plus some added hair.

Above we have given sufficient conditions for
\begin{equation}dist_C(n,n';G_0)=dist_C(n,n';G)\end{equation}
with $G$ an extension of $G_0$ and $n,n'\in V_0$. We show now that the
emergence of too short new paths is representing the obstruction for
such a result to hold in general.

So let again $G_0$ be a graph and assume the existence of two nodes,
$n,n'$, in $V_0$ with
\begin{equation}dist_C(n,n';G_0)>l\in\N\end{equation}
We extend $G_0$ to some $G$ by adding new nodes and edges. We know
that for admissible functions the elementary jumps along an edge have
to fulfill $|f_i-f_k|\leq 1$. If there exists a new path, $\gamma$, in $G$,
connecting $n,n'$, with
\begin{equation}length(\gamma)\leq l\end{equation}
we can conclude that for each admissible function on $G$ it must hold
\begin{equation}|f(n)-f(n')|\leq l\end{equation}
We hence have
\begin{lemma}If two nodes in $G_0$ have $dist_C(n,n';G_0)>l$ and if
  there exists a path, $\gamma$, in $G$, connecting $n,n'$ and having
  length $l\leq l$, it
  necessarily holds that
\begin{equation}dist_C(n,n';G)\leq l<dist_C(n,n';G_0)\end{equation}
\end{lemma}

Up to now we have derived upper bounds on $dist_C$ relative to
$dist_C$ on subgraphs or the canonical graph distance $d(n,n')$. In
the following we will derive a quite efficient lower bound. This is
done by defining a particular admissible function, depending on an
arbitrary base node, $n_0$. That is, we fix an arbitrary node, dubbed
$n_0$, and take as admissible function the canonical distance, divided
by the local vertex degree:
\begin{equation}f_{n_0}(n):=(v_n)^{-1/2}\cdot
  d(n_0,n)\;,\;f_{n_0}(n):=(v^{(out)}_n)^{-1/2}\cdot
  d(n_0,n)\;,\;f_{n_0}(n_0)=0\end{equation}
for undirected, directed graphs, respectively.

From our general results we have
\begin{equation}\|df\|=\sup_i\left(\sum_{j=1}^{v_i}|f_{k_j}-f_i|^2\right)^{1/2}\end{equation}
or $v_i$ replaced by $v_i^{(out)}$. Inserting the above particular
function we get
\begin{equation}\|df\|\leq 1\end{equation}
as each term, $|f_{k_j}-f_i|$, is either zero or one (depending of
whether the distance to the base point remains constant or changes by
$\pm 1$).
\begin{lemma}The functions, $f_{n_0}(n)$, $n_0$ an arbitrary node in
  $G$, are admissible.
\end{lemma}
With $n,n'$ two arbitrary nodes in $G$ we take $n$ as base point,
$n_0$, and have
\begin{equation}f_{n_0}(n')-f_{n_0}(n)=f_{n_0}(n')=(v_{n'})^{-1/2}\cdot
  d(n',n)\end{equation}
(as $f_{n_0}(n)=f_{n_0}(n_0)=0$). As $dist_C$ is the supremum over
admissible functions, we get:
\begin{satz}\label{Lower}
\begin{equation}dist_C(n,n')\geq(v_{(n,n')})^{-1/2}\cdot
  d(n,n')\end{equation}
with $v_{(n,n')}$ the minimum of the (out-) vertex degrees at $n,n'$
respectively.
\end{satz}
Note that one can of course either choose $n$ or $n'$ as base point in
the definition of the above admissible function.

\subsection{Examples}
The general results derived above should be compared with the
results in e.g. \cite{5} to \cite{7}. Choosing the symmetric
difference operator as ``Dirac operator'' in the case of the
one-dimensional lattice the authors in \cite{5,6} got a distance which
is \tit{strictly greater} than the ordinary distance but their choice
does not fulfill the above natural constraint given in Theorem
\ref{norm}. Note in particular that our operator $d$ is a map from
node- to bond-functions which is not the case in these examples.
In \cite{7} the authors employed a symmetric doubling of the
non-symmetric adjacency matrix of the one-dimensional directed
lattice, $\Z^1$ as Dirac operator. With $v_i^{(out)}=1$ in this
example, our above general estimate yields
\begin{equation}d(n,n')\leq dist_C(n,n')\leq d(n,n')\Rightarrow
  dist_C(n,n')=d(n,n')\end{equation}
that is, we get the same result for our Dirac operator as for the
choice made in \cite{7}.

We want to close this paper with the discussion of several examples
which show that, in general, it is quite a non-trivial task to
calculate $dist_C$. The first one is a simple warm-up exercise, the
second one is the one-dimensional non-directed lattice, $\Z^1$,
discussed also by some of the authors mentioned above (treated however
within their own schemes) and is not so simple. The last one is the
directed $\Z^2$-lattice, which we do not solve in closed form, but we
provide several estimates.

The technique used in approaching some of the problems may be
interesting in general.  It turns out that the proper mathematical
context, to which our strategy does belong, is the field of
\tit{(non-)linear programming} or \tit{optimization} (see e.g.
\cite{Jungnickel} or any other related textbook). This can be inferred
from the structure of the constraints we get. This means that the
techniques developed in this field may perhaps be of use in solving
such intricate problems.
\\[0.5cm]
\tit{Example 1: The square with vertices and edges}:
\begin{equation} x_1-x_2-x_3-x_4-x_1\end{equation}
Let us calculate the Connes-distance between $x_1$ and $x_3$.  As the
$\sup$ is taken over functions, the summation over elementary jumps is
(or rather: has to be) path-independent (this represents a subtle
constraint for practical calculations). It is an easy exercise to see
that the $sup$ can be found in the class where the two paths between
$x_1,x_3$ have the {\it valuations} ($1\geq a\geq0$):
\begin{equation} x_1-x_2:\,a\;,\;x_2-x_3:\,(1-a^2)^{1/2}\end{equation}
\begin{equation} x_1-x_4:\,(1-a^2)^{1/2}\;,\;x_4-x_3:\,a\end{equation}
Hence one has to find $\sup_{0\leq a\leq1}(a+\sqrt{1-a^2})$. Setting
the derivative with respect to $a$ to zero one gets
$a=\sqrt{1/2}$. That is:\\[0.3cm]
\begin{bsp}[Connes-distance on a square]
\begin{equation} dist_C(x_1,x_3)=\sqrt{2}<2=d(x_1,x_3)\end{equation}
\end{bsp}
\begin{bem}As $v_i=2$, our apriori estimate in theorem \ref{Lower} is
  saturated as
\begin{equation}dist_C(x_1,x_2)\geq (2)^{-1/2}\cdot
  2=\sqrt{2}\end{equation}

\end{bem}
The next example is considerably more complicated.
\\[0.5cm]\tit{Example 2: The undirected one-dimensional lattice}:\\[0.3cm]
The nodes are numbered by $\Z$. We want to calculate $dist_C(0,n)$
within our general framework. The calculation will be done in two main
steps. In the first part we make the (in principle quite complicated)
optimization process more accessible. For the sake of brevity we state
without proof that it is sufficient to discuss real monotonely
increasing functions with
\begin{equation}f(k)=\begin{cases}f(0) & \text{for}\;k\leq 0\\f(n) &
    \text{for}\;k\geq n \end{cases}
\end{equation}
and we write
\begin{equation}f(k)=f(0)+\sum_{i=1}^k h_i\quad\text{for}\quad 0\leq
  k\leq n\;h_i\geq 0\end{equation}
The above optimization process then reads:
\begin{ob}Find $\sup \sum_{i=1}^n h_i$ under the constraint
\begin{equation}h_1^2\leq 1,h_2^2+h_1^2\leq
  1,\ldots,h_n^2+h_{n-1}^2\leq 1,h_n^2\leq 1\end{equation}
\end{ob}
The simplifying idea is now the following. Let $h:=(h_i)_{i=1}^n$ be
an admissible sequence with \tit{all} $h_{i+1}^2+h_i^2<1$. We can then
find another admissible sequence $h'$ with
\begin{equation}\sum h_i'>\sum h_i\end{equation}
Hence the supremum cannot be taken on the interior. We conclude that
at least some $h_{i+1}^2+h_i^2$ have to be one. There is then a
minimal $i$ for which this holds. We can convince ourselves that the
process can now be repeated for the substring ending at $i+1$.
Repeating the argument we can fill up all the entries up to place
$i+1$ with the condition $h_{l+1}^2+h_l^2=1$ and proceeding now
upwards we end up with
\begin{lemma}The above supremum is assumed within the subset
\begin{equation}h_1^2\leq
  1,h_1^2+h_2^2=1,\ldots,h_{n-1}^2+h_n^2=1,h_n^2\leq 1\end{equation}
\end{lemma}
This concludes the first step.

In the second step we calculate $\sup|f(0)-f(n)|$ on this restricted
set. From the above we now have the constraint:
\begin{equation}h_1^2\leq
  1,h_2^2=1-h_1^2,h_3^2=h_1^2,h_4^2=1-h_1^2,\ldots,h_n^2=1-h_1^2\;\text{or}\;h_1^2 \end{equation}
depending on $n$ being even or uneven. This yields
\begin{equation}\sup|f(0)-f(n)|=\begin{cases}
1 & \text{for}\;n=1\\
(n/2)\cdot\sup(h_1+\sqrt{1-h_1^2})=(n/2)\cdot\sqrt{2} &
\text{for $n$ even}\\
\sup ({[}n/2]\cdot(h_1+\sqrt{1-h_1^2})+h_1) & \text{for $n$ uneven}
\end{cases}
\end{equation}
In the even case the rhs can be written as
$\sqrt{n^2/2}=\sqrt{[n^2/2]}$. In the uneven case we get by
differentiating the rhs and setting it to zero:
\begin{equation}h_1^{max}=A_n/\sqrt{1+A_n^2}\;,\;\sqrt{1-(h_1^{max})^2}=1/\sqrt{1+A_n^2}\end{equation}
with $A_n=1+1/[n/2]$. We see that for increasing $n$ both terms
approach $1/\sqrt{2}$, the result in the even case. Furthermore we see that the
distance is monotonely increasing with $n$ as should be the case for a
distance. This yields in the uneven case
\begin{equation}\label{uneven} dist_C(0,n)=\frac{([n/2]+1)A_n+[n/2]}{\sqrt{1+A_n^2}}\end{equation}
which is a little bit nasty. Both expressions can however be written in a
more elegant and unified way (this was a conjecture by W.Kunhardt,
inferred from numerical examples). For $n$ uneven a short calculation yields
\begin{equation}[n^2/2]=(n^2-1)/2=1/2\cdot(n-1)(n+1)=2[n/2]([n/2]+1)\end{equation}
(with the \tit{floor-,ceiling-}notation the expressions would become
even more elegant). With the help of the latter formula the rhs in
(\ref{uneven}) can be transformed into
\begin{equation}\frac{([n/2]+1)A_n+[n/2]}{\sqrt{1+A_n^2}} =\sqrt{[n^2/2]+1}\end{equation}
\begin{conclusion}For the one-dimensional undirected lattice we have
\begin{equation}dist_C(0,n)=\begin{cases} \sqrt{[n^2/2]} & \text{for
      $n$ even}\\\sqrt{[n^2/2]+1} & \text{for $n$ uneven}
\end{cases}
\end{equation}
\end{conclusion}
\begin{bem}Again comparing the exact result with our lower bound, we
  find for $n$ even:
\begin{equation}dist_C(0,n)\geq (2)^{-1/2}\cdot n=(2)^{1/2}\cdot n/2\end{equation}
that is, the lower bound is again saturated. For $n$ uneven we have
instead:
\begin{equation}dist_C(0,n)=(2)^{-1/2}\cdot(n^2+1)^{1/2}>(2)^{-1/2}\cdot
  n\end{equation}
\end{bem}
\tit{Example 3: The directed lattice $\Z^2$}\\[0.3cm]
The vertices in $\Z_d^2$ are denoted by $(i,j)$ or $(x,y)$. The edges
point from $(i,j)$ to $(i+1,j)$ and $(i,j+1)$; hence,
$v_i^{out}=2$. As the system is translation invariant, it suffices to
calculate the Connes distance between nodes $(0,0)$ and $(x,y)$ with
$x,y>0$.

For nodes lying on the same parallel to the $x$-, $y$-axis,
respectively, we have
\begin{equation}dist_C(n,n')=d(n,n')\end{equation}
For $x$ or $y=0$, there is only one minimal path, connecting $(0,0)$
and $(x,y)$. Therefore, lemma \ref{equal} does not apply. For, say,
$y=0$, we choose the following \tit{admissible} function:
\begin{equation}f(x,y):=x\quad\text{for all}\quad y\end{equation}
We have
\begin{equation}|f(x,0)-f(0,0)|=|x|=d((0,0),(x,0))\end{equation}
and can conclude
\begin{equation}dist_C((0,0),(x,0))=d((0,0),(x,0))\end{equation}
on $\Z^2_d$. The same holds for the $y$-axis.

For nodes with both $x,y\neq 0$, we have more than one minimal path
connecting $(0,0)$ and $(x,y)$. Our lemma then shows that,
necessarily,
\begin{equation}dist_C(n,n')<d(n,n')\end{equation}
More detailed estimates will be given below.

If we try to really calculate the Connes distance on $\Z^2_d$ for
points in \tit{general position}, the optimization problem becomes
quite involved and we will only provide some estimates. The reason is
that the constraint equations are of a quite \tit{non-local} nature
(compared to the simpler undirected $\Z^1$-lattice) and that in
general several minimal paths do exist which make the continuation
problem quite intricate.

It is easy to see that the canonical graph distance between the points
$(0,0)$ and (x,y) is $|x|+|y|$ and that all minimal paths have the
same length. With $x,y>0$, we conjecture (without giving a proof) that it
suffices to restrict the variation to admissible functions with
positive jumps in the positive $x-,y-$directions and that we can set
$f(0,0)=0$. A particular minimal path consists of $x$ steps in the
$x$-direction followed by $y$ steps in the $y$-direction. We denote
(for convenience) the jumps along the $x-,y-$axis, respectively, by
\begin{equation}h_{i0}:=f(i,0)-f(i-1,0)\geq
  0\;h_{0j}:=f(0,j)-f(0,j-1)\geq 0\end{equation}
The optimization problem now reads:
\begin{problem}Find
  $\sup\left(\sum_{i=1}^xh_{i0}+\sum_{j=1}^yh_{0j}\right)$ under the
  constraints imposed by the admissibility of the corresponding
  function, $f$. Note however that the constraints must hold on the
  full lattice.
\end{problem}

From our general result in theorem \ref{Lower} we know that
\begin{equation}dist_C((0,0),(x,y))\geq
  (2)^{-1/2}\cdot(x+y)\end{equation}
We can construct an admissible function, $f$, which fulfills
\begin{equation}|f(0,0)-f(x,y)|=(2)^{-1/2}\cdot(x+y)\end{equation}
This can be achieved by setting $f(0,0)=0$ and by choosing all
$x-,y-$jumps equal to $a$ with $2a^2=1\Rightarrow a=(2)^{-1/2}$. This
yields the above result.

The question is, whether this is already the supremum over the set of
admissible functions. We will show, that this is not the case by
providing an other admissible function yielding a bigger value. We
choose an admissible function with $x$-jumps equal to $a$ and
$y$-jumps equal to $b$ with $a^2+b^2=1$. The admissible function reads
\begin{equation}f(x,y)=ax+by\end{equation}
The function $f$ takes a \tit{stationary value} at
\begin{equation}a=(x^2/(x^2+y^2))^{1/2}\;b=(y^2/(x^2+y^2))^{1/2}\end{equation}yielding
the value
\begin{equation}f(x,y)=(x^2+y^2)^{1/2}\end{equation}
Assuming, for example, that $x\neq y$, it follows that\begin{equation}
f(x,y)^2-((2)^{-1/2}\cdot(x+y))^2=(1/2)\cdot(x^2+y^2)-x\cdot
y=(1/2)\cdot(x-y)^2>0\end{equation}
In other words, we see
\begin{ob}For the directed $\Z^2$-lattice and $x\neq y$ ($x,y>0$) we have the
  estimate
\begin{equation}d((0,0),(x,y))>dist_C((0,0),(x,y))\geq
  (x^2+y^2)^{1/2}>(2)^{-1/2}(x+y)\end{equation}
\end{ob}

\end{document}